\documentclass[aps,prd,amsmath,amssymb]{revtex4}

\usepackage{graphicx}
\usepackage{dcolumn}
\usepackage{bm}
\usepackage{amsmath}
\usepackage{epstopdf}
\usepackage{amsfonts}
\usepackage{amssymb}
\usepackage{epsfig}
\usepackage{tabularx}
\usepackage{color}
\usepackage{soul}

\usepackage[colorlinks=true,citecolor=blue,urlcolor=magenta,breaklinks]{hyperref}

\newcommand{\be}{\begin{equation}}
\newcommand{\en}{\end{equation}}
\newcommand{\bea}{\begin{eqnarray}}
\newcommand{\ena}{\end{eqnarray}}

\begin{document}

\title{Interacting dark sector: Lagrangian formulation based on two canonical scalar fields}

\author{Grigoris Panotopoulos}  \email[]{grigorios.panotopoulos@ufrontera.cl}
\affiliation{Centro de Astrof{\'i}sica e Gravita{\c c}{\~a}o-CENTRA, Instituto Superior T{\'e}cnico-IST, Universidade de Lisboa-UL, Av. Rovisco Pais, 1049-001 Lisboa, Portugal,}
\affiliation{Departamento de Ciencias F{\'i}sicas, Universidad de la Frontera, Avenida Francisco Salazar, 01145 Temuco, Chile.}

\author{Il{\'i}dio Lopes}  \email[]{ilidio.lopes@tecnico.ulisboa.pt}
\affiliation{Centro de Astrof{\'i}sica e Gravita{\c c}{\~a}o-CENTRA, Instituto Superior T{\'e}cnico-IST, Universidade de Lisboa-UL, Av. Rovisco Pais, 1049-001 Lisboa, Portugal.}

\begin{abstract}
We study in some detail an interacting cosmological model based on two canonical scalar fields  starting from a Lagrangian description. Contrary to other more phenomenological approaches where non-relativistic matter and dark energy are cosmological perfect fluids, and where a source term is added by hand at the level of the continuity equations, here within Einstein's theory we model the dark sector, which dominates the evolution of the universe, as two minimally coupled scalar fields, out of which the first play the role of dark matter and the second plays the role of dark energy. We compute both the deceleration parameter and the distance modulus versus red-shift, and we demonstrate that the model is capable of explaining the current cosmic acceleration. We find that a negative coupling constant implies two distinctive features, 
which comprise two robust predictions of the model studied in the present work, and which are the following: a) a transient acceleration phase, and b) a collapsing Universe, or in other words an initial expansion which is followed by a contraction leading eventually to a Big-Crunch.
\end{abstract}

\maketitle

\section{Introduction}

Over the last 25 years or so, thanks to advances in modern technology and sophisticated space-based detectors and telescopes, we have entered an era of precision Cosmology. Currently, a plethora of astrophysical and cosmological probes  provide us with strong evidence that we live in a spatially flat Universe, which undergoes an accelerating expansion dominated by dark matter and dark energy \cite{MTurner}. Unveiling the origin and nature of the dark sector comprises one 
of the major challenges in modern theoretical cosmology. Einstein's General Relativity (GR), despite its mathematical beauty and successful confrontation with a series of experimental and observational data \cite{CWill}, cannot provide us with accelerating solutions if the energy content of the universe includes radiation and non-relativistic matter only. The simplest and most economical model capable of explaining the current cosmic acceleration is the concordance cosmological model -- Lambda cold dark matter ($\Lambda$CDM), which is based on cold dark matter combined with a positive and tiny cosmological constant \cite{SCarroll}, and which overall is in a very good agreement with a great deal of current observational data. Cosmological constant may be viewed as a perfect fluid with an equation-of-state parameter $w=p/\rho=-1$, with $p,\rho$ being the pressure and energy density of the fluid, respectively. Since, however, it suffers from the cosmological constant problem \cite{SWeinberg} as well as the "why now" (or coincidence) problem, other alternatives have been proposed and studied over the years. A dynamical dark energy \cite{ECopeland} characterized by an evolving equation-of-state parameter, $w(a)$, with $a$ being the scale factor, interacting with cold dark matter may alleviate the aforementioned two problems related to the cosmological constant, and therefore interacting dark energy models have attracted a lot of attention over the years. 
For an incomplete list see e.g. \cite{model1,model2,model3,model4,model5,model6,model7} and references therein. 

Although a phenomenological approach based on different dark energy parameterizations \cite{Leandros,Cardenas} is quite simple as it is straightforward to obtain an analytic expression of the Hubble parameter as a function of the red-shift, $H(z)$, a more fundamental description based on a Lagrangian formulation is desirable and more advantageous. The simplest choice is to introduce scalar fields, since they carry no indices and therefore i) they are compatible with isotropy and homogeneity of the Universe, and ii) their study is less complicated from the mathematical point of view. After the discovery of the Standard Model Higgs boson almost 10 years ago \cite{ATLAS,CMS}, we know that fundamental scalars do exist in nature. What is more, scalar fields arise in many different contexts in modern particle physics. Some well-known examples are the following: i) Higgs bosons required to break electroweak symmetry, and give masses to 
particles \cite{higgs1,higgs2}, ii) pseudo-Goldstone bosons associated with explicit breaking of additional global symmetries \cite{freese}, iii) moduli coming from Superstring theory compactifications \cite{moduli1,moduli2,moduli3,moduli4,moduli5}, iv) scalar fields included in supermultiplets in supersymmetric models \cite{SMartin} and theories of supergravity \cite{HNilles}, to mention just a few.

In the present work we propose to study an interacting dark energy model within GR starting from a Lagrangian formulation based on two canonical scalars fields, where one of the fields plays the role of dark matter and the other plays the role of dark energy. It is well known after \cite{coherent} that coherent oscillations of a scalar field around the minimum of a monomial potential of the form $\phi^2$ acts like a pressure-less fluid, i.e., dust. This is the approach followed in a similar work several years ago \cite{OBertolami}. There is, however, another way to reproduce a matter-dominated era, and this is based on power-law solutions for the scale factor using a scalar potential of exponential form \cite{powerlaw}. In the present work we shall follow the second approach. The interaction term between the two scalar field, too, is different
than the one used in \cite{OBertolami}. Here we adopt the simplest potential compatible with renormalizability and a discrete $Z_2$ parity.

The plan of work is the following: In the next section we set the scene presenting briefly the cosmological equations we shall be using throughout this work including scalar field Cosmology based on a single minimally coupled scalar field. In the third section we introduce the interacting dark sector model analyzed here, and in the Section IV we present and discuss our numerical results. Finally, we finish with some concluding remarks in Section IV. We adopt the mostly positive metric signature $-,+,+,+$, and we work in geometrized units where $c=1=8 \pi G$.

\section{Theoretical framework}

\subsection{Basic cosmological equations}

The starting point is Einstein's GR \cite{GR} based on the Einstein-Hilbert term coupled to the matter content
\begin{equation}
S = \int d^4x \sqrt{-g} \left[ \frac{R}{2}  + \mathcal{L}_M \right]
\end{equation}
where $g_{\mu \nu}$ is the metric tensor, $g$ is its determinant, $R$ is the corresponding Ricci scalar, $G$ is Newton's constant and $\mathcal{L}_M$ is the Lagrangian of the matter content. Varying the action with respect to the metric tensor
one obtains the well-known Einstein's field equations, which read
\begin{equation}
G_{\mu \nu} \equiv R_{\mu \nu} - \frac{1}{2} R g_{\mu \nu} =  T_{\mu \nu}
\end{equation}
where $R_{\mu \nu}$ is the Ricci tensor, while $T_{\mu \nu}$ is the matter energy-momentum tensor.

The basic cosmological equations governing the expansion of a homogeneous and isotropic Universe may be found e.g. in
\cite{review}. If matter consists of a perfect fluid with pressure $p$ and energy density $\rho$, the energy momentum 
tensor is given by
\begin{equation}
T_{\mu \nu} = p \: g_{\mu \nu} + (p + \rho) \: u_\mu \: u_\nu 
\end{equation}
where $u_\mu$ is the four-velocity of the fluid satisfying the condition $u_\mu u^\mu = -1$. The mixed component
stress-energy tensor takes the form \cite{review}
\begin{equation}
T_{\nu}^{\mu} =\textrm{diag}(-\rho,p,p,p)
\end{equation}

An isotropic and homogeneous universe (spatially flat $k=0$) is described by a Robertson-Walker metric \cite{review}
\begin{equation}
ds^2 = -dt^2 + a(t)^2 \delta_{ij} dx^i dx^j 
\end{equation}
where the scale factor $a(t)$ is the only unknown quantity, and all quantities depend on the cosmic time $t$ only.

The cosmological equations are found to be the continuity equation as well as the two Friedmann
equations \cite{review}
\begin{eqnarray}
H^2 & = & \frac{1}{3} \rho \\
\frac{\ddot{a}}{a} & = & -\frac{1}{6} (\rho + 3p) \\
0 & = & \dot{\rho} + 3 H (\rho+p)
\end{eqnarray}
where an over dot denotes differentiation with respect to cosmic time, and $H=\dot{a}/a$ is the Hubble parameter. The
second Friedmann equation may be written down equivalently as follows
\begin{equation}
\dot{H} = \frac{\ddot{a}}{a} - H^2 = - \frac{\rho+p}{2}
\end{equation}
If there are several non-interacting fluid components, then
\begin{eqnarray}
p & = & \sum_i p_i \\
\rho & = & \sum_i \rho_i \\
0 & = & \dot{\rho_i} + 3 H (\rho_i+p_i)
\end{eqnarray}
For barotropic fluids $p=w \rho$, where $w$ is the equation-of-state parameter.

Let us now focus on an expanding Universe dominated by dark energy and dark matter, allowing for an interaction term 
between the two fluid components. The Friedmann equations remain the same, but the continuity equations take the form
\begin{eqnarray}
-Q & = & \dot{\rho}_m + 3 H \rho_m \\
Q & = & \dot{\rho}_{\chi}+ 3 H \rho_{\chi}(1+w)
\end{eqnarray}
where $Q$ is the source term describing the energy flow between dark matter and dark energy, and the equation-of-state for dark matter $w_m=0$. A simple and concrete interacting model in which analytic expressions may be obtained is the following: Assuming that the source term is proportional to the matter energy density, $Q=\delta H \rho_m$, 
with $\delta$ being the coupling of the interaction, the fluid equations take the form
\begin{eqnarray}
-\delta H \rho_m & = & \dot{\rho}_m + 3 H \rho_m \\
\delta H \rho_m & = & \dot{\rho}_{\chi}+ 3 H \rho_{\chi}(1+w)
\end{eqnarray}
In the following we shall assume for simplicity that both $w$ and $\delta$ are constants. In this case it is 
straightforward to integrate the fluid equations, and obtain $\rho_m,\rho_{\chi}$ in terms of the scale factor $a$.
Finally, introducing the red-shift, $1+z = a_0/a$, with $a_0$ being the present value of the scale factor, the
Hubble parameter as a function of the red-shift is computed to be \cite{interacting}
\begin{equation}
E(z)^2 \equiv (H(z)/H_0)^2 = \Omega_{\chi}(1+z)^{3 (1+w)} + \frac{1-\Omega_{\chi}}{\delta+3w} [\delta (1+z)^{3 (1+w)}+3 w (1+z)^{3-\delta}]
\end{equation} 
The deceleration parameter, $q$, is defined to be
\begin{equation}
q \equiv - \frac{\ddot{a}}{a H^2}
\end{equation}
and as a function of red-shift it is  given by
\begin{equation}
q(z) = -1 + (1+z) \frac{H'(z)}{H(z)}.
\end{equation}

\subsection{Scalar field Cosmology}

A scalar field slowly rolling down its potential (quintessence) \cite{Quint,cosmon} acts as an effective cosmological constant, and tracker \cite{trackers} or scaling solutions \cite{scaling} may be obtained to address the coincidence problem. In scalar field Cosmology, the starting point is a Lagrangian containing the usual Einstein-Hilbert term coupled to a canonical (minimally coupled) scalar field, $\phi$, with a kinetic term and a potential term $V(\phi)$. The Lagrangian of the scalar field has the form
\begin{equation}
\mathcal{L}_M = -\frac{1}{2} (\partial \phi)^2 - V(\phi) \equiv K - V(\phi)
\end{equation}
The stress-energy tensor of the scalar field is computed to be
\begin{equation}
T_{\mu \nu} = \partial_\mu \phi \: \partial_\mu \phi + \mathcal{L}_M \: g_{\mu \nu}
\end{equation}
and it can be written down equivalently as the energy-momentum tensor of a perfect fluid provided that
\begin{eqnarray}
\rho_\phi & = & K + V(\phi) \\
p_{\phi} & = & K - V(\phi) \\
u_\mu & = & \frac{\partial_\mu \phi}{\sqrt{2K}}
\end{eqnarray}
while for a FRW Universe they are simply given by
\begin{eqnarray}
\rho_\phi & = & \frac{1}{2} \dot{\phi}^2 + V(\phi) \\
p_{\phi} & = & \frac{1}{2} \dot{\phi}^2 - V(\phi)
\end{eqnarray}
Therefore the continuity equation for a fluid is equivalent to the Klein-Gordon equation
\begin{equation}
\ddot{\phi} + 3 H \dot{\phi} + V_{,\phi} = 0
\end{equation}
while the two Friedmann equations now take the form
\begin{eqnarray}
H^2 & = & \frac{1}{3} \left( \frac{1}{2} \dot{\phi}^2 + V(\phi) \right) \\
\dot{H} & = &  - \frac{\dot{\phi}^2}{2}
\end{eqnarray}
It is not difficult to show that a power-law solution for the scale factor, $a(t) \propto t^s$, is an exact analytic solution to the cosmological equations, provided that the scalar potential takes the exponential form \cite{powerlaw}
\begin{equation}
V(\phi) = V_0 exp(-\lambda \phi)
\end{equation}
where the exponent, $\lambda$, is directly related to the power $s$ as follows
\begin{equation}
\lambda = \sqrt{2 / s}.
\end{equation}
Any power $s > 1$ corresponds to accelerating solutions, whereas $s < 1$ corresponds to decelerating solutions. In particular, matter dominated era with $s=2/3$ may be reproduced in scalar field Cosmology adopting a scalar potential of exponential form with exponent $\lambda=\sqrt{3}$.

\section{Interacting dark sector: Lagrangian formulation}

\subsection{System of coupled equations}

Finally, in this section we introduce a model for the dark sector where a non-vanishing interaction between dark energy and dark matter is allowed. This possibility may be realized introducing two canonical scalar fields, $\phi,\chi$, with a total scalar potential of the form
\begin{equation}
W(\phi,\chi) = V_1(\phi) + V_2(\chi) + V_{int}(\phi,\chi)
\end{equation}
where the first term is the scalar potential of the first scalar field alone, the second term of the second scalar field alone, while the last term is responsible for the interaction between the two scalar fields. Defining for each field a momentum as follows \cite{RKallosh}
\begin{eqnarray}
P_\phi & \equiv & a^3 \dot{\phi} \\
P_\chi & \equiv & a^3 \dot{\chi}
\end{eqnarray}
the complete system of coupled equations may be written down as follows \cite{RKallosh}
\begin{eqnarray}
\dot{a} & = & a H \\
\dot{\phi} & = & \frac{P_\phi}{a^3} \\
\dot{\chi} & = & \frac{P_\chi}{a^3} \\
\dot{P_\phi} & = & -a^3 W_{,\phi} \\
\dot{P_\chi} & = & -a^3 W_{,\chi} \\
\dot{H} & = & -H^2 + \frac{P_\phi^2 + P_\chi^2}{3 a^6} - \frac{W}{3}
\end{eqnarray}
where in the numerical analysis we have considered the case
\begin{eqnarray}
V_i(z) & = & V_{0,i} \exp{\left[- \lambda_i z \right]} \\
V_{int}(\phi,\chi) & = & \frac{g}{4} \phi^2 \: \chi^2
\end{eqnarray}
with $g$ being a dimensionless coupling constant, which is essentially the strength of the interaction. The interacting potential is the simplest one compatible with renormalizability and a $Z_2$ symmetry, $\phi \rightarrow -\phi$ and 
$\chi \rightarrow -\chi$, or in other words it treats both fields on equal footing. Since we follow the evolution of the
Universe starting from the matter era, we impose for the scale factor the initial condition \cite{NegPot}
\begin{equation} 
a(t_i) \sim (9 \Omega_{m,0}/3)^{1/3} t_i^{2/3}
\end{equation}
while for the scalar fields we assume the following initial conditions
\begin{eqnarray}
\phi(t_i) & = & \phi_i \\
\dot{\phi}(t_i) & = & 0 \\
\chi(t_i) & = & \chi_i \\
\dot{\chi}(t_i) & = & 0 
\end{eqnarray}
Finally, the first Friedmann equation is used as a constraint to fix the initial condition for the Hubble parameter.

\subsection{Numerical results}

Once the solution to the system of coupled cosmological equations is obtained, a number of useful quantities may be computed. First, the dark energy equation-of-state, $w_\phi$, by definition is given by
\begin{equation} 
w_\phi = \frac{p_\phi}{\rho_\phi} = \frac{\frac{1}{2} \dot{\phi}^2 - V_1(\phi)}{\frac{1}{2} \dot{\phi}^2 + V_1(\phi)}.
\end{equation}
The equation-of-state parameter of a canonical scalar field evolves with time remaining always in the range $-1 < w_\phi < 1$. This may be easily seen considering two limiting cases where the kinetic term, $\dot{\phi}^2/2$, is much larger or much lower than the potential term, $V_1(\phi)$, and the dark energy equation-of-state $w_\phi$ reduces to 1 or -1, i.e.
\begin{equation} 
w_\phi \approx -1, \, \, \, \, \, \, \dot{\phi}^2/2 \ll V_1(\phi)
\end{equation}
\begin{equation} 
w_\phi \approx 1, \, \, \, \, \, \, \dot{\phi}^2/2 \gg V_1(\phi)
\end{equation}

Moreover, the distance modulus $\mu=m-M$, where $m$ and $M$ are the apparent and absolute magnitude, respectively, is given by \cite{Leandros,hogg}
\begin{equation} 
\mu(z) = 25 + 5 \: \log_{10} \left[ \frac{D_L(z)}{\text{Mpc}} \right]
\end{equation}
where the luminosity distance, $D_L(z)$, is given by \cite{Leandros,hogg}
\begin{equation}
D_L(z) = (1+z) \int_0^z dx \frac{1}{H(x)}.
\end{equation}

\begin{figure*}[ht!]
\centering
\includegraphics[width=0.55\textwidth]{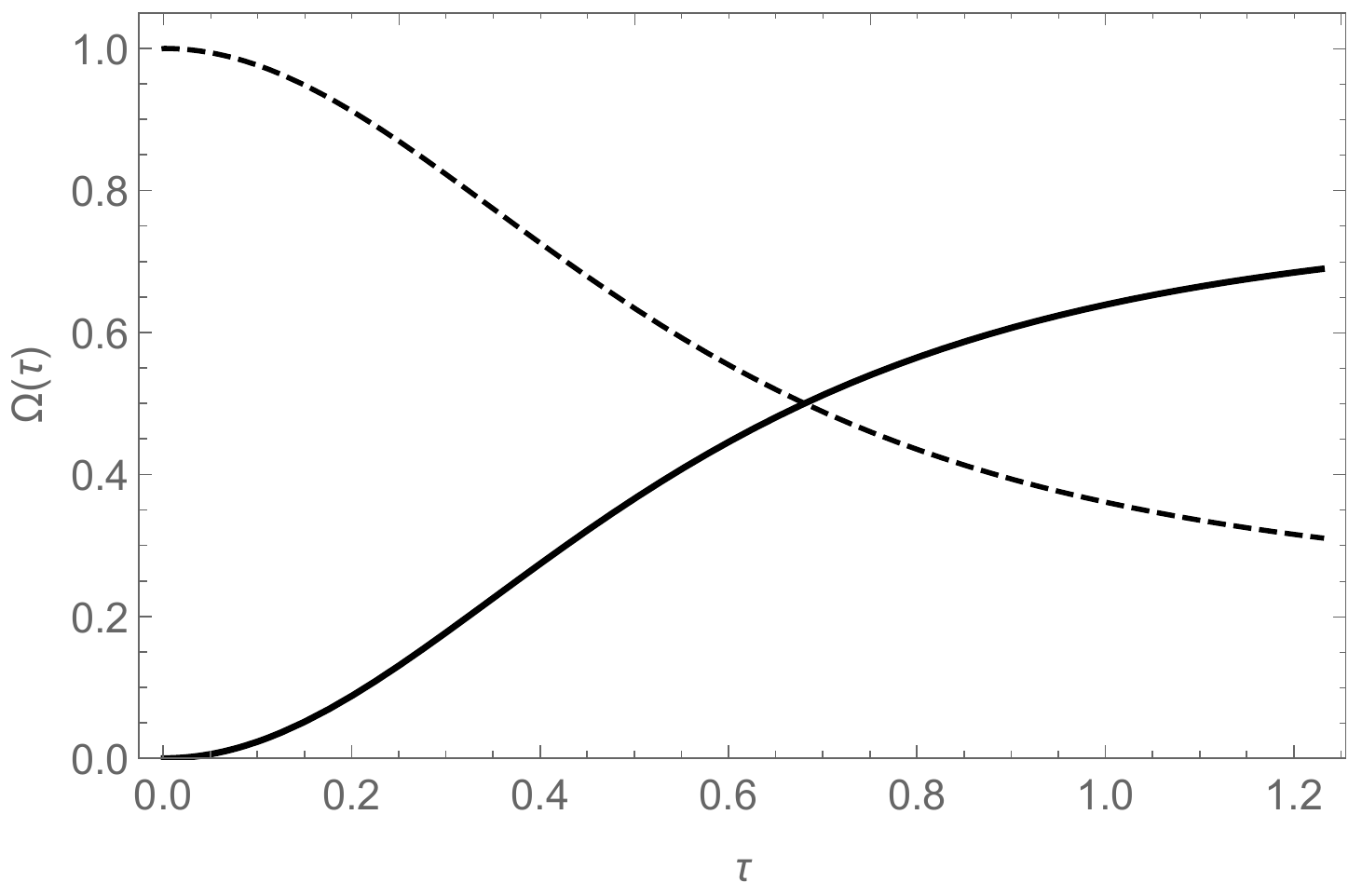}
\caption{ 
Dark sector modeled introducing two canonical scalar fields. The figure shows the evolution of the normalized densities of dark energy (solid) and non-relativistic matter (dashed) versus cosmic time for a non-interacting model, $g=0$.
}
\label{fig:Figure1}
\end{figure*}

Our main numerical results are summarized in the figures below. It is demonstrated that the model studied here exhibits
the expected behavior of any viable dark energy model. Although we have not performed a detailed comparison between the predictions of the model and current data, at this level of discussion it is shown that the model at least meets the minimum
requirements. In particular, in Fig.~\ref{fig:Figure1} we show the evolution of the normalized densities as a function of the cosmic time. Initially the evolution of the Universe is dominated by matter, whereas later on dark energy started to dominate the expansion of the Universe. Finally, today the normalized densities of dark energy and matter acquire the values $0.7,0.3$, respectively.

\begin{figure*}[ht!]
\centering
\includegraphics[width=0.48\textwidth]{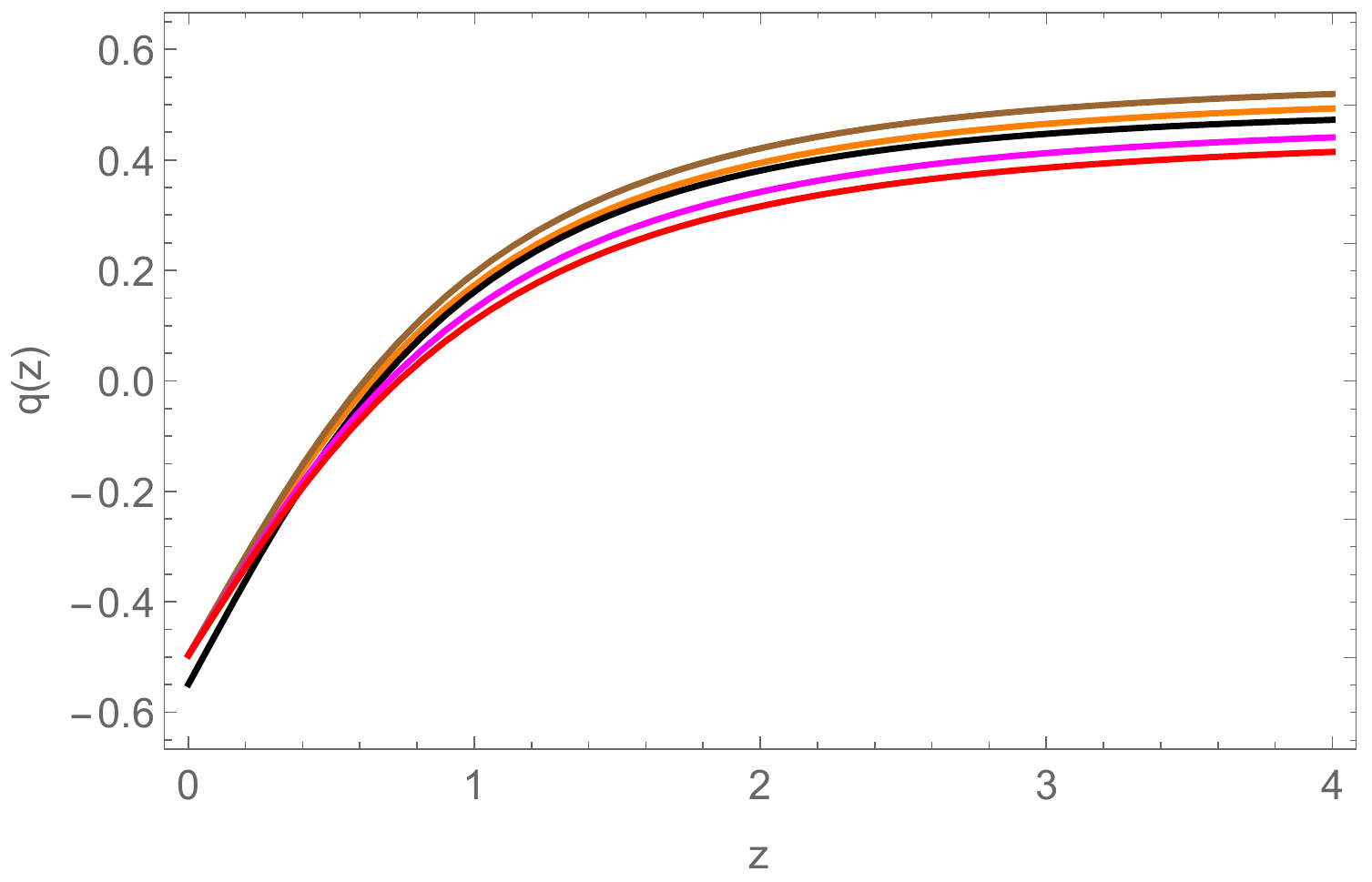}   \\
\includegraphics[width=0.48\textwidth]{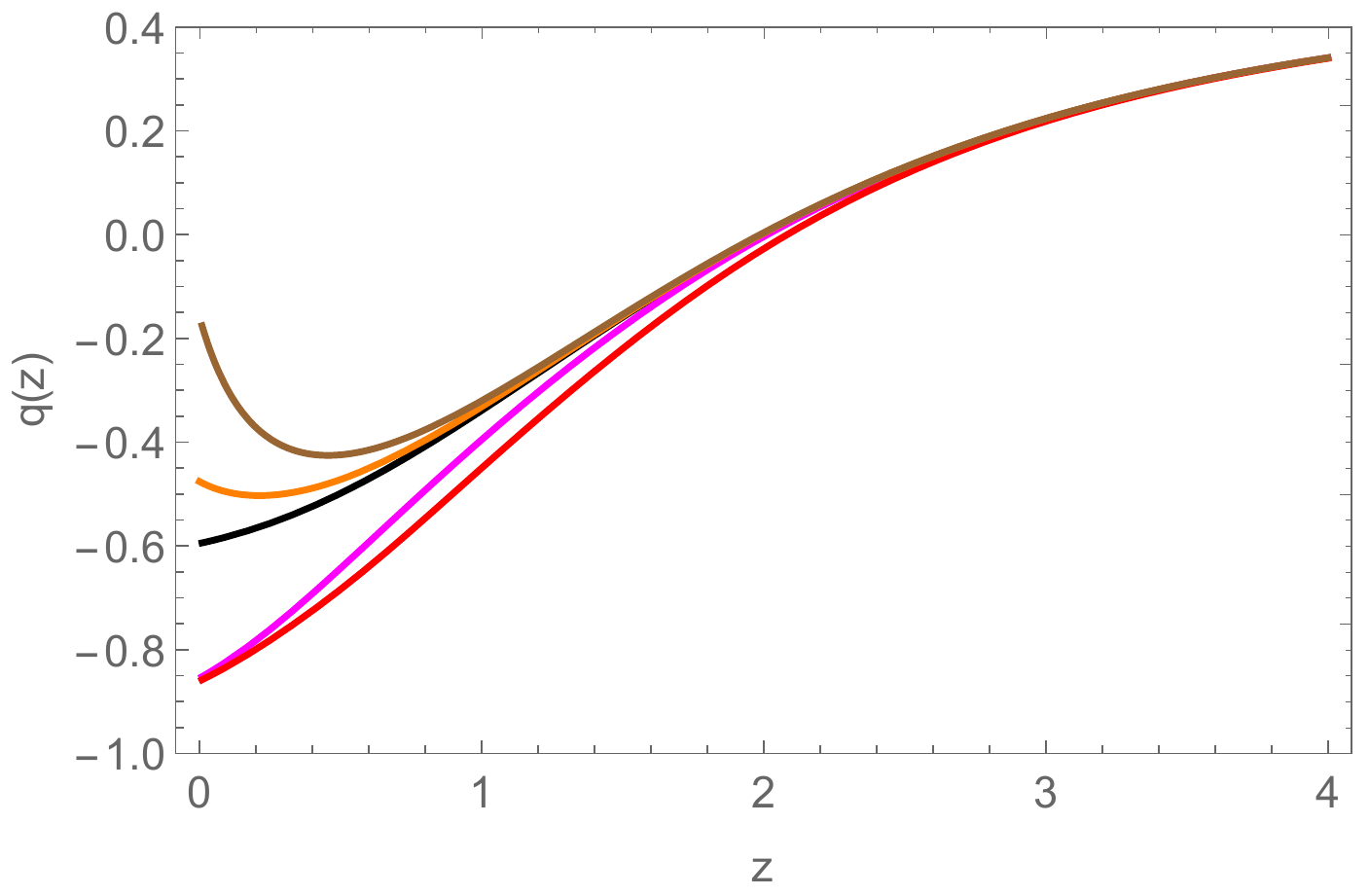} 
\caption{ 
Deceleration parameter $q$ versus red-shift $z$ for positive (red and magenta), negative (orange and brown) and 
zero (black) coupling constants. 
{\bf TOP:} Case of phenomenological description based on fluid components. 
{\bf BOTTOM:} Case of Lagrangian formulation based on canonical scalar fields. 
}
\label{fig:Figure2}
\end{figure*}

Next, in Fig.~\ref{fig:Figure2} we show the deceleration parameter, $q$, versus red-shift, $z$, for positive (red and magenta curves), negative (orange and brown curves) and zero (black curve) coupling constant $g $ and $\delta$. The upper panel corresponds to the phenomenological description based on two perfect fluids, whereas the lower panel corresponds to the Lagrangian formulation based on two canonical scalar fields. The following features are observed: a) in both cases the Universe passes from a decelerating era to an accelerated era at late times, b) when the interacting dark sector is described introducing scalar fields the decelerating parameter changes its sign at somewhat higher red-shift, c) in the case of the Lagrangian formulation a negative coupling constant slightly slows down the current acceleration, and d) in the fluid description of the dark sector, all curves become indistinguishable during the accelerating phase, $q<0$, whereas in the Lagrangian formulation the opposite holds. Future measurements of the present value $q_0$ may discriminate between several different dark energy models.

\begin{figure*}[ht!]
\centering
\includegraphics[width=0.52\textwidth]{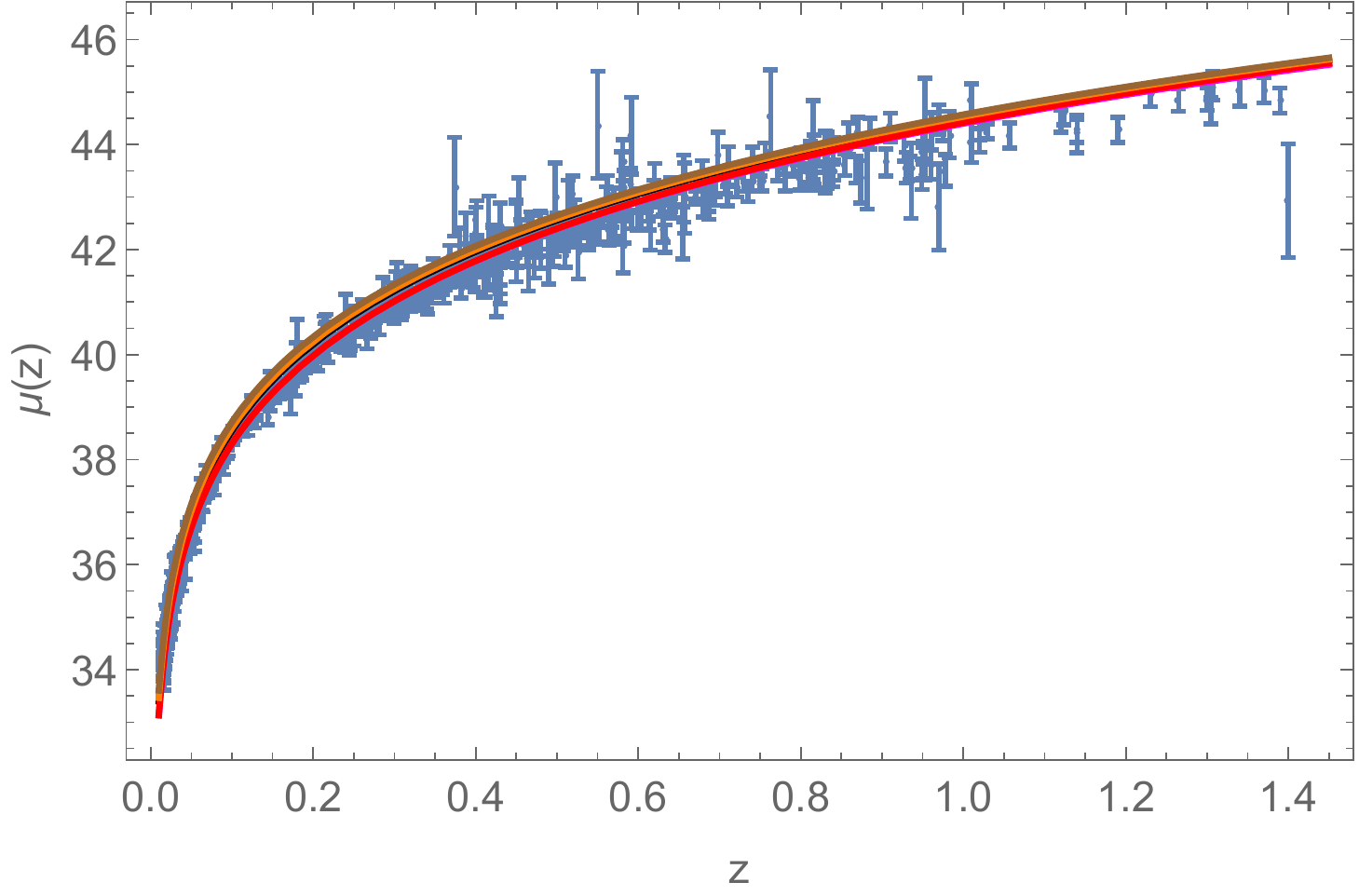}   \\
\includegraphics[width=0.55\textwidth]{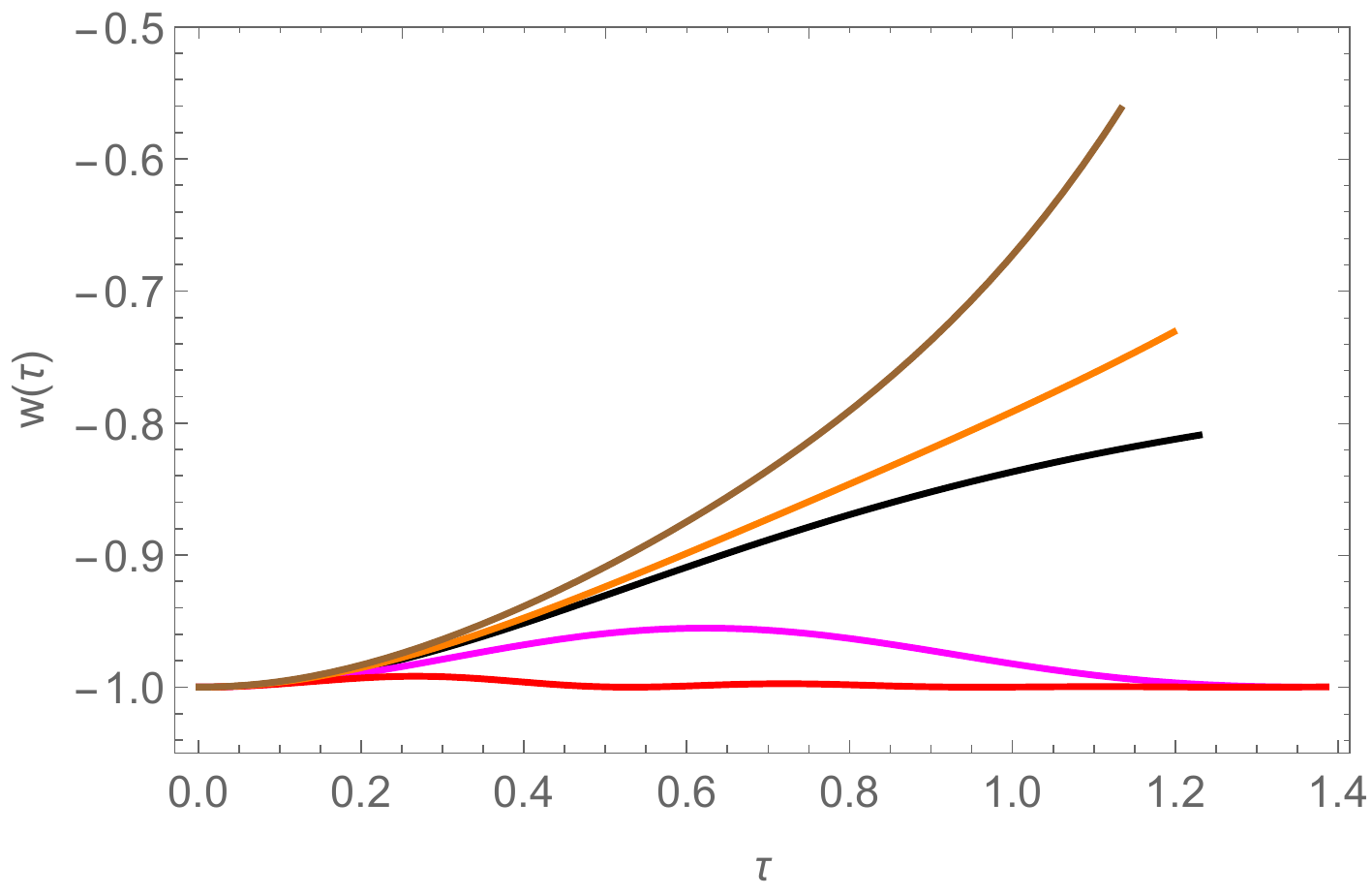} 
\caption{ 
Dark energy models based on two canonical scalar fields.
{\bf TOP:} Distance modulus $\mu$ versus red-shift $z$. The data of the Union2 compilation are shown as well. 
Shown are: $g=0$ (in black),  $g > 0$ (in red and magenta) and $g < 0$ (in orange and brown).
{\bf BOTTOM:} Dark energy equation-of-state parameter, $w_{\chi}=p_\phi/\rho_\phi$, versus dimensionless time,
$\tau=H_0 t$ for $g=0$ (in black),  $g > 0$ (in red and magenta) and $g < 0$ (in orange and brown).
}
\label{fig:Figure3}
\end{figure*}

Furthermore, in Fig.~\ref{fig:Figure3} we show the distance modulus, $\mu$, versus red-shift, $z$ (upper panel), as well as the dark energy equation-of-state, $w_{\chi}$ (lower panel), in the case of the Lagrangian formulation based on two canonical scalar fields for zero (black curve), positive (cyan curve) and negative (red curve) coupling constant $g$. The supernovae data points from the Union2 compilation \cite{union} are shown as well for comparison reasons. Clearly, there is a very good agreement between the models considered here and the data points.

\begin{figure*}[ht!]
\centering
\includegraphics[width=0.5\textwidth]{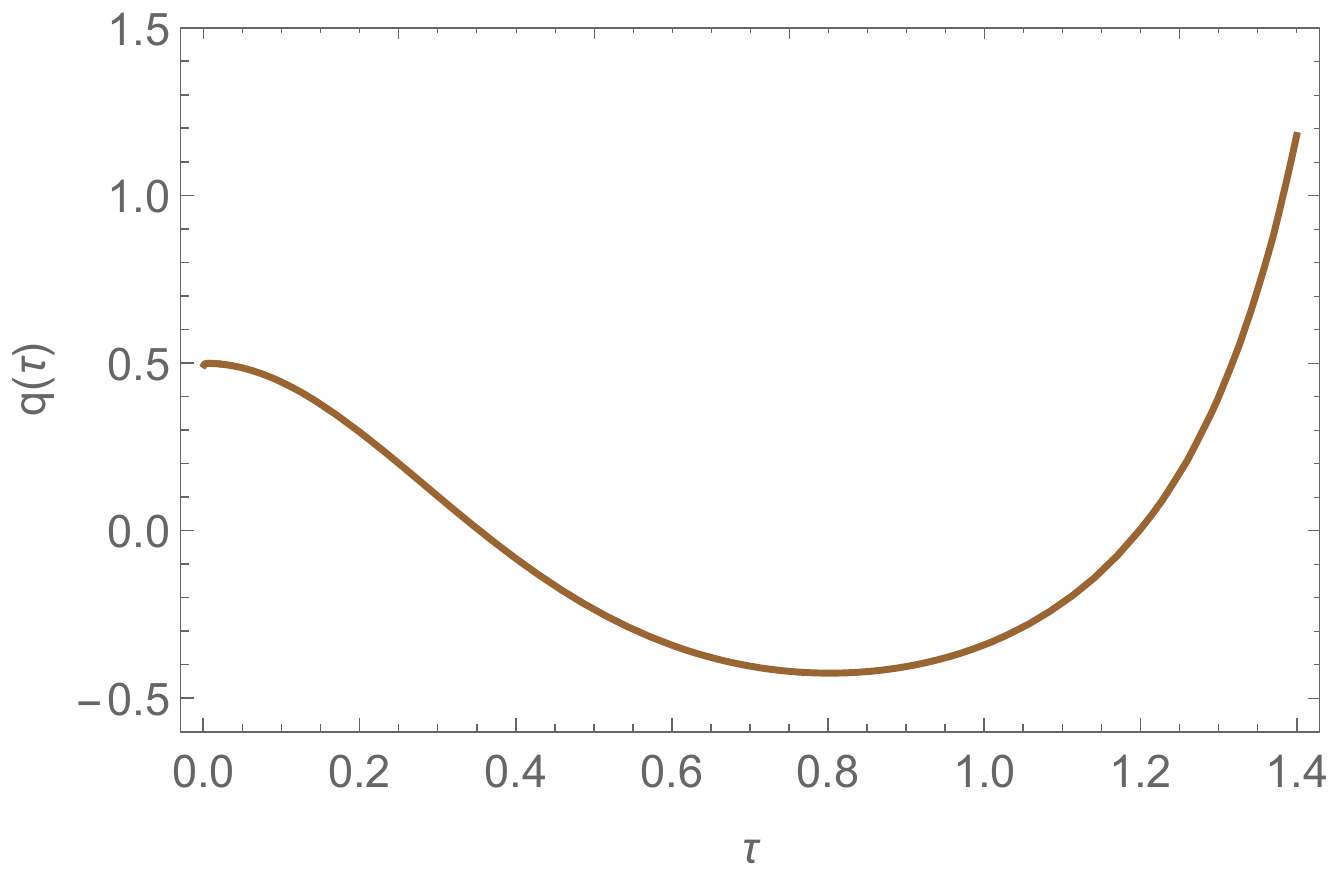}   \\
\includegraphics[width=0.5\textwidth]{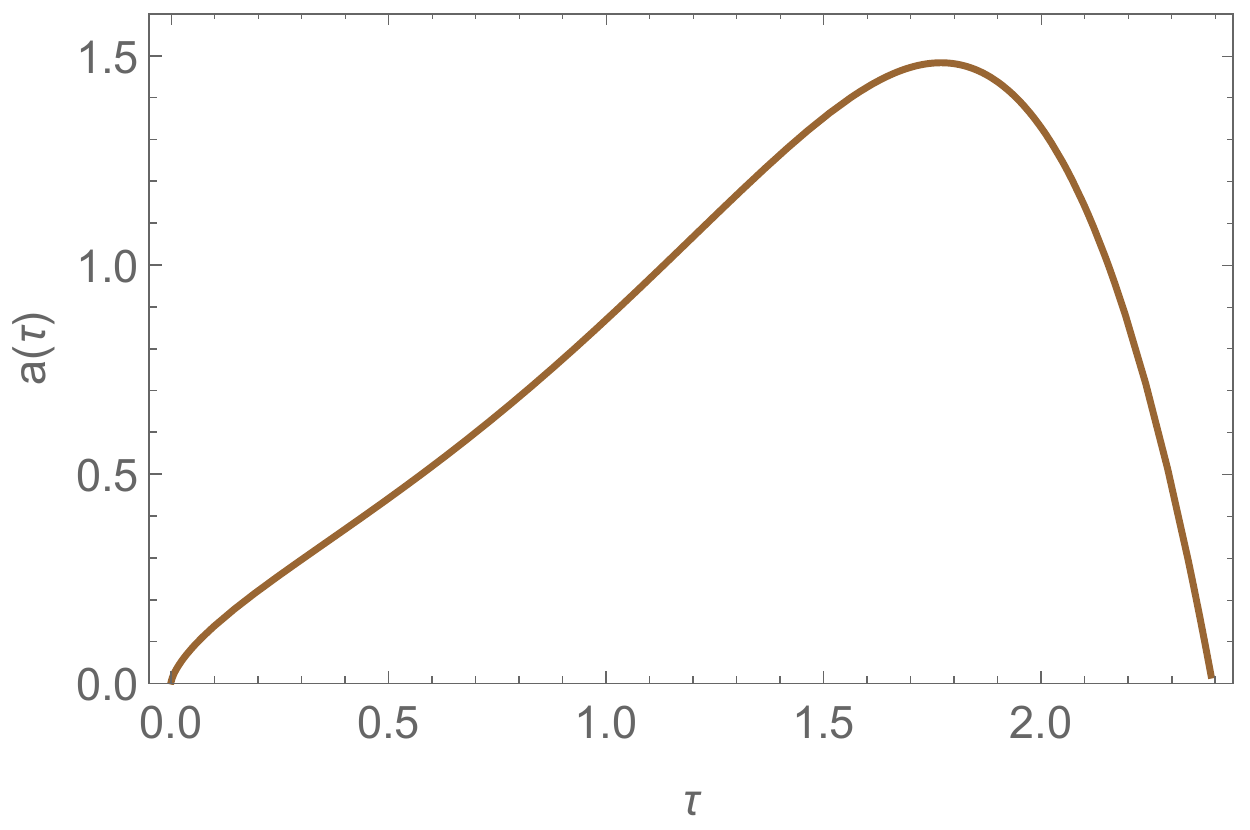} 
\caption{ 
Dark energy models based on two canonical scalar fields and negative coupling constant. 
{\bf TOP:} An expanding Universe exhibiting a transient acceleration phase between two decelerating phases. 
{\bf BOTTOM:} Scale factor versus dimensionless time $\tau=H_0 t$ for a collapsing Universe. The Big-Crunch occurs 
at $t_* = 2.11 \: t_0$.
}
\label{fig:Figure4}
\end{figure*}

Finally, let us comment on a couple of interesting features (shown in the last figure) observed when the coupling constant is negative. The top panel of Fig.~\ref{fig:Figure4} shows the deceleration parameter as a function of the dimensionless time, $\tau=H_0 t$, with $H_0$ being the Hubble constant, $H_0=H(t_0)$, and the age of the Universe is determined by $a(t_0)=1$.
The current cosmic acceleration is slowing down, and in the future the Universe eventually enters a second decelerating phase. What is more, the lower panel of the same figure shows the scale factor versus cosmic time for a collapsing Universe,
where the expansion is followed by a contraction leading finally to a Big-Crunch at $t_* = 2.1 \: t_0$. This possibility 
for the fate of the Universe depending on the nature of the dark energy model has been discussed in \cite{fate1,fate2}.
A transient acceleration phase was observed in \cite{OBertolami} as well, although nothing regarding a collapsing Universe was mentioned there.


Before we summarize our work, a final remark is in order. A non-vanishing interaction between matter and dark energy
violates weak equivalence principle, and it modifies the well-known virial theorem, see e.g. \cite{bertolami1,bertolami2}. In the non-relativistic limit, if $K$ is the Newtonian kinetic energy, and $W$ is the gravitational potential energy, then the modified virial theorem, irrespectively of the interaction model, takes the form \cite{bertolami1}
\begin{equation}
2 \rho_K + \rho_W = \zeta \rho_W
\end{equation}
where $\zeta$ is the coupling constant of the matter-DE interaction, while the quantities $\rho_K, \rho_W$ are defined 
to be
\begin{equation}
\rho_K \equiv M \frac{dK}{dV}, \; \; \; \; \; \rho_W \equiv M \frac{dW}{dV}.
\end{equation}
Clearly, when there is no interaction between matter and dark energy one obtains the usual virial ratio, $\rho_K/\rho_W=-0.5$.

It turns out that suitable gravitationally bounded astronomical objects may be used to probe the validity of the above expression. A relaxed cluster, characterized by an approximately spherically symmetric mass profile, is in fact the Abell Cluster A586 \cite{Abell}. For that particular distribution, the densities $\rho_K, \rho_W$ are computed to be \cite{bertolami1}
\begin{equation}
\rho_K = \frac{9 M \sigma^2}{8 \pi R^3}, \; \; \; \; \; \rho_W = - \frac{3 G M^2}{8 \pi R^3 \langle R \rangle},
\end{equation}
where $M$ and $R$ are the total mass and the radius of the object, respectively, while $\sigma$ is the dispersion velocity, and $\langle R \rangle$ is the mean intergalactic distance \cite{Abell}.

In particular for the Abell Cluster those quantities are computed to be \cite{Abell}
\begin{eqnarray}
M & = & 4.3 \times 10^{14}~M_{\odot} \\
\sigma & = & 1243~km/s \\
\langle R \rangle & = & 309~kpc
\end{eqnarray}
with $M_{\odot}$ being the solar mass, and therefore the coupling constant $\zeta$ is found to be
\begin{equation}
-0.774 = \frac{\rho_K}{\rho_W} = \frac{\zeta-1}{2}, \; \; \; \; \; \zeta=-0.548.
\end{equation}
Therefore, the Abell Cluster A586 provides strong evidence for a non-vanishing, in fact negative, coupling
constant between the components of the dark sector.

\section{Conclusions}

In summary, we have studied within Einstein's General Relativity an interacting dark energy model based on a Lagrangian formulation. Two canonical (minimally coupled to gravity) scalar fields, $\phi,\chi$, have been introduced, which play the role of dark energy and dark matter, respectively. Each scalar field comes with its own self-interaction potential of exponential form, while at the same time we have included an interaction potential of the form $g \phi^2 \chi^2$, characterized by a dimensionless coupling constant $g$, that can be either positive or negative. with two properties, namely a) it is of renormalizable type, and b) it treats both fields on equal footing, or in other words it has a $Z_2$ parity,
$\phi \rightarrow -\phi$, $\chi \rightarrow -\chi$. We have written down the cosmological equations as a system of first order differential equations, where the second Friedmann equations is a dynamical equations, whereas the first Friedmann equation is a constraint. The system of equations has been integrated numerically, and the effect of the sign of the coupling constant is investigated in some detail. Our numerical results indicate that a negative coupling constant is more interesting. It has been demonstrated that the model is capable of describing the cosmic current acceleration, and that it is in a very good agreement with available supernovae data. In particular, we have computed i) the deceleration parameter as a function of the red-shift, where a decelerating phase is followed by an accelerating one, and ii) the distance modulus as a function of red-shift, where the data from the Union2 compilation are shown as well. Our main numerical results may be summarized as follows: i) the passage from deceleration to acceleration takes place at higher red-shift compared to the concordance $\Lambda$CDM model (at least for the numerical values of the parameters considered here), ii) a negative coupling constant is more interesting, since on the one hand seems to be slowing down the current acceleration, and on the other hand it leads to a Big-Crunch (contraction followed by an expansion).

\section*{Acknowlegements}

We thank the anonymous referee for suggestions. The authors G.~P. and I.~L. thank the 
Funda\c c\~ao para a Ci\^encia e Tecnologia (FCT), Portugal, for the financial support 
to the Center for Astrophysics and Gravitation-CENTRA, Instituto Superior T\'ecnico, 
Universidade de Lisboa, through the Project No.~UIDB/00099/2020 and grant No. PTDC/FIS-AST/28920/2017.




\begin{thebibliography}{99}

\bibitem{MTurner}  W.~L.~Freedman and M.~S.~Turner,
  Rev.\ Mod.\ Phys.\  {\bf 75} (2003) 1433
  [astro-ph/0308418].
  
\bibitem{CWill} C.~M.~Will,
  Living Rev.\ Rel.\  {\bf 17} (2014) 4
  [arXiv:1403.7377 [gr-qc]].  

\bibitem{SCarroll} S.~M.~Carroll,
  Living Rev.\ Rel.\  {\bf 4} (2001) 1
  [astro-ph/0004075].

\bibitem{SWeinberg} S.~Weinberg,
  Rev.\ Mod.\ Phys.\  {\bf 61} (1989) 1.

\bibitem{ECopeland} E.~J.~Copeland, M.~Sami and S.~Tsujikawa,
  Int.\ J.\ Mod.\ Phys.\ D {\bf 15} (2006) 1753
  [hep-th/0603057].

\bibitem{model1} L.~Amendola, G.~Camargo Campos and R.~Rosenfeld,
  Phys.\ Rev.\ D {\bf 75} (2007) 083506
  [astro-ph/0610806].
  
\bibitem{model2} V.~Salvatelli, N.~Said, M.~Bruni, A.~Melchiorri and D.~Wands,
  Phys.\ Rev.\ Lett.\  {\bf 113} (2014) no.18,  181301
  [arXiv:1406.7297 [astro-ph.CO]].

\bibitem{model3} S.~Das, P.~S.~Corasaniti and J.~Khoury,
  Phys.\ Rev.\ D {\bf 73} (2006) 083509
  [astro-ph/0510628].

\bibitem{model4} G.~Kofinas, G.~Panotopoulos and T.~N.~Tomaras,
  JHEP {\bf 0601} (2006) 107
  [hep-th/0510207].

\bibitem{model5} L.~Lopez Honorez, O.~Mena and G.~Panotopoulos,
  Phys.\ Rev.\ D {\bf 82} (2010) 123525
  [arXiv:1009.5263 [astro-ph.CO]].

\bibitem{model6} C.~G.~Boehmer, G.~Caldera-Cabral, N.~Chan, R.~Lazkoz and R.~Maartens,
  Phys.\ Rev.\ D {\bf 81} (2010) 083003
  [arXiv:0911.3089 [gr-qc]].
  
 \bibitem{model7} V.~H.~Cardenas and S.~Lepe,
  Eur.\ Phys.\ J.\ C {\bf 80} (2020) no.9,  862
  [arXiv:2008.13577 [astro-ph.CO]].

\bibitem{Leandros} S.~Nesseris and L.~Perivolaropoulos,
  Phys.\ Rev.\ D {\bf 70} (2004) 043531
  [astro-ph/0401556].

\bibitem{Cardenas} J.~Maga{\~n}a, V.~H.~C{\'a}rdenas and V.~Motta,
  JCAP {\bf 1410} (2014) 017
  [arXiv:1407.1632 [astro-ph.CO]].

\bibitem{ATLAS} G.~Aad \textit{et al.} [ATLAS],
Phys. Lett. B \textbf{716} (2012), 1-29
[arXiv:1207.7214 [hep-ex]].

\bibitem{CMS} S.~Chatrchyan \textit{et al.} [CMS],
Phys. Lett. B \textbf{716} (2012), 30-61
[arXiv:1207.7235 [hep-ex]].

\bibitem{higgs1} P.~W.~Higgs,
  Phys.\ Rev.\ Lett.\  {\bf 13} (1964) 508.

\bibitem{higgs2} F.~Englert and R.~Brout,
  Phys.\ Rev.\ Lett.\  {\bf 13} (1964) 321.

\bibitem{freese} A.~Dolgov and K.~Freese,
  Phys.\ Rev.\ D {\bf 51} (1995) 2693
  [hep-ph/9410346].
  
\bibitem{moduli1} X.~de la Ossa and E.~E.~Svanes,
  JHEP {\bf 1410} (2014) 123
  [arXiv:1402.1725 [hep-th]].
  
\bibitem{moduli2} J.~Louis,
  NATO Sci.\ Ser.\ C {\bf 556} (2000) 61.

\bibitem{moduli3} E.~E.~Svanes,
  arXiv:1411.6696 [hep-th].
  
\bibitem{moduli4} R.~Galvez,
  Phys.\ Rev.\ D {\bf 94} (2016) no.10,  103521
  [arXiv:1603.06631 [hep-th]].
  
\bibitem{moduli5} J.~Gray and H.~Parsian,
  JHEP {\bf 1807} (2018) 158
  [arXiv:1803.08176 [hep-th]].

\bibitem{SMartin} S.~P.~Martin,
  Adv.\ Ser.\ Direct.\ High Energy Phys.\  {\bf 21} (2010) 1
   [Adv.\ Ser.\ Direct.\ High Energy Phys.\  {\bf 18} (1998) 1]
  [hep-ph/9709356].

\bibitem{HNilles} H.~P.~Nilles,
  Phys.\ Rept.\  {\bf 110} (1984) 1.  
  
\bibitem{coherent} M.~S.~Turner,
Phys. Rev. D \textbf{28} (1983), 1243.

\bibitem{OBertolami} O.~Bertolami, P.~Carrilho and J.~Paramos,
  Phys.\ Rev.\ D {\bf 86} (2012) 103522
  [arXiv:1206.2589 [gr-qc]].

\bibitem{powerlaw} F.~Lucchin and S.~Matarrese,
  Phys.\ Rev.\ D {\bf 32} (1985) 1316.  

\bibitem{GR} A.~Einstein,
  Annalen Phys.\  {\bf 49} (1916) no.7,  769
  [Annalen Phys.\  {\bf 354} (1916) no.7,  769].

\bibitem{review} D.~Huterer and D.~L.~Shafer,
  Rept.\ Prog.\ Phys.\  {\bf 81} (2018) no.1,  016901
  [arXiv:1709.01091 [astro-ph.CO]].

\bibitem{interacting} Z.~K.~Guo, N.~Ohta and S.~Tsujikawa,
Phys. Rev. D \textbf{76} (2007), 023508
[arXiv:astro-ph/0702015 [astro-ph]].

\bibitem{Quint} B.~Ratra and P.~J.~E.~Peebles,
  Phys.\ Rev.\ D {\bf 37} (1988) 3406.
  
\bibitem{cosmon} C.~Wetterich,
  Astron.\ Astrophys.\  {\bf 301} (1995) 321
  [hep-th/9408025]. 
  
\bibitem{trackers} P.~J.~Steinhardt, L.~M.~Wang and I.~Zlatev,
  Phys.\ Rev.\ D {\bf 59} (1999) 123504
  [astro-ph/9812313].
  
\bibitem{scaling}  A.~R.~Liddle and R.~J.~Scherrer,
  Phys.\ Rev.\ D {\bf 59} (1999) 023509
  [astro-ph/9809272].  
  
\bibitem{RKallosh} R.~Kallosh and S.~Prokushkin,
  hep-th/0403060.
  
\bibitem{NegPot} L.~Perivolaropoulos,
  Phys.\ Rev.\ D {\bf 71} (2005) 063503
  [astro-ph/0412308].  
  
\bibitem{hogg} D.~W.~Hogg,
  astro-ph/9905116.  

\bibitem{union} N.~Suzuki {\it et al.},
Astrophys.\ J.\  {\bf 746} (2012) 85
  [arXiv:1105.3470 [astro-ph.CO]].
  
\bibitem{fate1} R.~Kallosh and A.~D.~Linde,
  JCAP {\bf 0302} (2003) 002
  [astro-ph/0301087].
  
\bibitem{fate2} R.~Kallosh, J.~Kratochvil, A.~D.~Linde, E.~V.~Linder and M.~Shmakova,
  JCAP {\bf 0310} (2003) 015
  [astro-ph/0307185].
  
\bibitem{bertolami1}O.~Bertolami, F.~Gil Pedro and M.~Le Delliou,
Phys. Lett. B \textbf{654} (2007), 165-169
[arXiv:astro-ph/0703462 [astro-ph]]. 

\bibitem{bertolami2} M.~Le Delliou, O.~Bertolami and F.~Gil Pedro,
AIP Conf. Proc. \textbf{957} (2007) no.1, 421-424
[arXiv:0709.2505 [astro-ph]].

\bibitem{Abell} E.~S.~Cypriano, G.~B.~Neto, L.~J.~Sodr{\'e} and J.~P.~Kneib,
Ap. J. {\bf 630} (2005) 38.


\end{thebibliography}
\end{document}